\documentclass[%
 reprint, amsmath,amssymb,
 aps,prl]{revtex4-1}

\usepackage{graphicx}
\usepackage{dcolumn}
\usepackage{bm}
\usepackage[colorlinks=true,linkcolor=blue]{hyperref}%

\begin{document}

\title{Emission of plasmons by drifting Dirac electrons: where hydrodynamics matters}

\author{Dmitry Svintsov}
 \email{svintcov.da@mipt.ru}
\affiliation{Laboratory of 2d Materials for Optoelectronics, Moscow Institute of Physics and Technology
}

\date{\today}

\begin{abstract}
Direct current in clean semiconductors and metals was recently shown to obey the laws of hydrodynamics in a broad range of temperatures and sample dimensions. However, the determination of frequency window for hydrodynamic phenomena remains challenging. Here, we reveal a phenomenon being a hallmark of high-frequency hydrodynamic transport, the Cerenkov emission of plasmons by drifting Dirac electrons. The effect appears in hydrodynamic regime only due to reduction of plasmon velocity by electron-electron collisions below the velocity of carrier drift. To characterize the Cerenkov effect quantitatively, we analytically find the high-frequency non-local conductivity of drifting Dirac electrons across the hydrodynamic-to-ballistic crossover. We find the growth rates of hydrodynamic plasmon instabilities in two experimentally relevant setups: parallel graphene layers and graphene covered by subwavelength grating, further showing their absence in ballistic regime. We argue that the possibility of Cerenkov emission is linked to singular structure of non-local conductivity of Dirac materials and is independent on specific dielectric environment.


\end{abstract}

\maketitle

The realm of hydrodynamic transport spans at length scales exceeding the particle free path~\cite{Physical_Kinetics}. Experimental demarcation of hydrodynamics and ballistics is conveniently performed by measuring the flow through a pipe between two reservoirs. The flow through wide pipes is limited by viscosity (Poiseuille flow), while in narrower pipes it is limited by particle injection (Knudsen flow). Recently, similar experiments were performed in ultra-clean solid-state systems, including thin metal wires~\cite{PdCo_hydro}, Weyl semimetals~\cite{WP2_Hydro}, GaAs-based quantum wells~\cite{Levin_NonlocalGaAs}, and graphene~\cite{Bandurin_Negative,Bandurin_superballistic,Poiseuille_flow}. They have revealed quite a broad window of temperatures and sample dimensions where electrons obey the laws of hydrodynamics~\cite{Adam_HD_Window} but not ballistics, as thought previously~\cite{Mayorov_ballistic}. 

While the place for dc hydrodynamic (HD) phenomena on temperature and length scales is established~\cite{Fluidity_onset,Levitov_negative_nonlocal}, the bounds for hydrodynamics on frequency scale are less probed~\cite{Gallagher_High-frequency}. Generally, electron-electron (e-e) collisions being the prerequisite of HD transport affect neither dc nor ac conductivity in {\it uniform} fields, though they may affect the properties of waves in solids -- plasmons. Still, the spectra of plasmons in HD and ballistic regimes are almost identical as they are dictated by long-range Coulomb forces insensitive to microscopic details of e-e interactions~\cite{BGK_model,Abrikosov_TFL}. The character of damping due to e-e scattering in ballistic and HD regimes is different~\cite{Vignale_elasticity,Ashcroft_conserving}, still, it is often masked by extrinsic damping.

In this Letter, we theoretically reveal a plasmonic phenomenon serving as a hallmark of hydrodynamic transport, and is fully prohibited in collisionless ballistic regime. The effect is emission of plasmons by drifting Dirac electrons or, in other words, Cerenkov plasmon instability of electron drift. Our emphasis on Dirac electron systems, especially graphene, is motivated by numerous observations of hydrodynamic phenomena therein~\cite{Bandurin_Negative,Bandurin_superballistic,Poiseuille_flow,Crossno}; though some fingerprints of effect can be found in systems of massive electrons.  

The possibility of Cerenkov instability in the HD regime is not merely due to reduction of viscous dissipation. It appears due to softening of plasmon velocity by e-e collisions down to the value sufficient to provide phase synchronism between drifting carriers and waves. More precisely, the lower bound on ballistic plasmon velocity in materials with Dirac spectra $\epsilon_p = \pm p v_0$ is exactly the carrier velocity $v_0$~\cite{ryzhii_graphene_plasmon,Lundeberg_Nonlocal}. The velocity of drift $u_0 < v_0$ thus never satisfies the Cerenkov criterion. In the HD regime, the lower bound on plasmon velocity is only $v_0/\sqrt{D}$, where $D$ is the dimension of space~\cite{Our-hydrodynamic,Crossover}. The carriers accelerated to drift velocity $u_0>v_0/\sqrt{D}$ are thus capable of plasmon emission.

Previous studies of current-driven plasmon emission in graphene were a field of delusions, with spurious instabilities in ballistic regime predicted~\cite{Gumbs_Instability,Neagtive_LD}. The inadequacy of these predictions stems from breakdown of Galilean invariance in Dirac systems~\cite{Levitov_plasmons} which makes the Doppler transform inapplicable for plasmon frequencies in a moving reference frame~\cite{Comment}. More accurate studies~\cite{VanDuppen_birefringent,Wenger_CurrentControlled,Sabbaghi_Drift-induced} revealed no Cerenkov-type instabilities, but were limited to the collisionless case~\footnote{The instability obtained in~\cite{Sabbaghi_Drift-induced} is related to current-induced interband population inversion which can be obtained near the charge neutrality point. Here, we consider only the intra-band plasmon emission.}. 

\begin{figure}[ht]
	\includegraphics[width=0.85\linewidth]{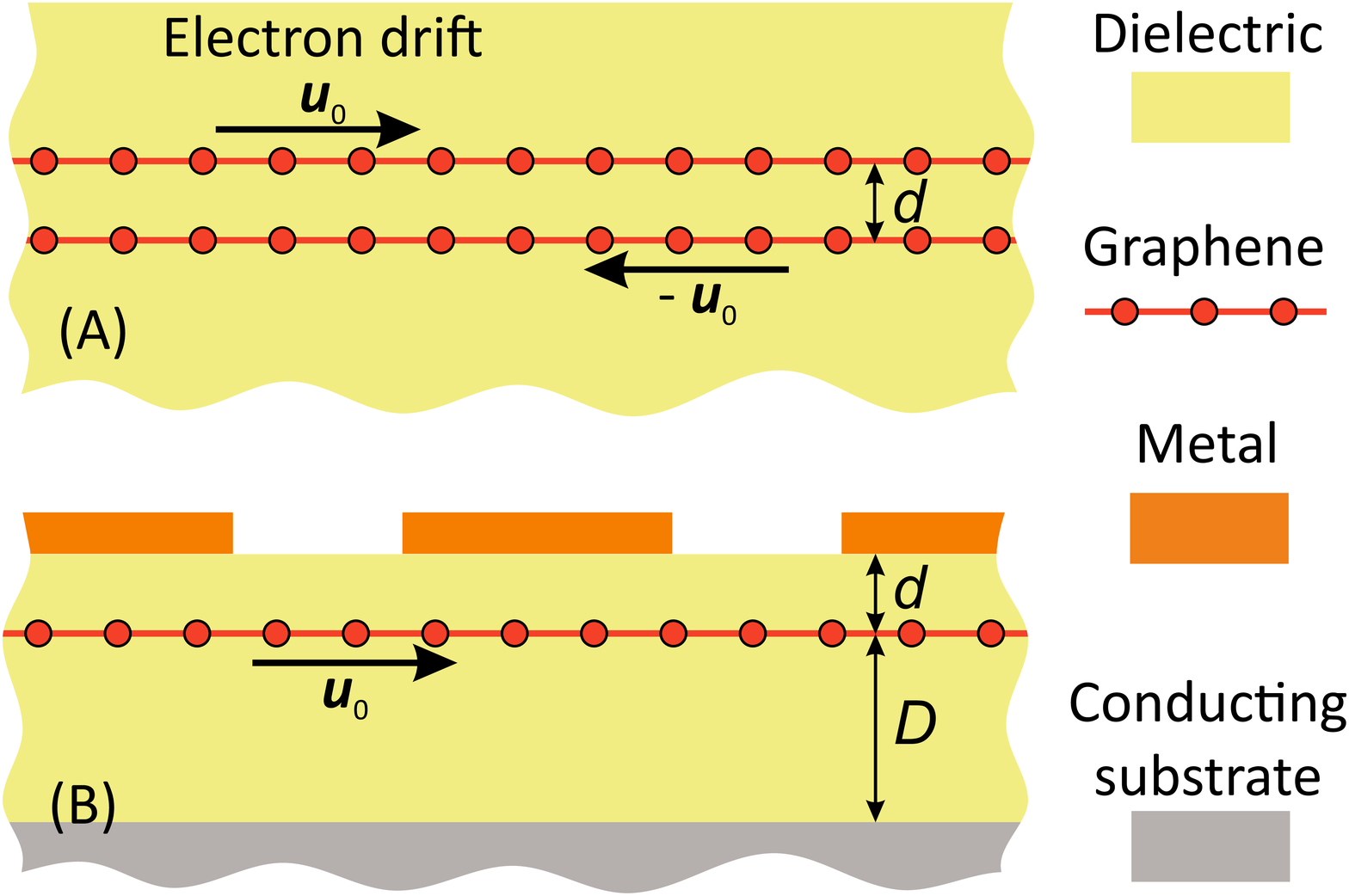}
	\caption{\label{Fig0} 
		Two possible graphene-based setups where hydrodynamic plasmon instabilities can be observed (A) parallel layers with counter-streaming electrons with velocities $\pm u_0$ (B) graphene covered with sub-wavelength plasmonic grating and a highly-conducting substrate}
\end{figure}

Below we construct the theory of plasmon instabilities in Dirac systems that can handle the subtle issues of Galilean invariance breakdown. Moreover, it is capable to trace the evolution of instabilities across the hydrodynamic-to-ballistic crossover {\it analytically}. It is based on solution of kinetic equation with model e-e collision integral satisfying the conservation laws. The obtained conductivity $\sigma(q,\omega)$ of drifting Dirac electrons has a number of unexpected features, including the absence of dissipation at special frequencies and wave vectors satisfying $\omega u_0 = qv_0^2$. It is subsequently used as a building block for analysis of current-driven plasmon instabilities in experimentally relevant setups, including parallel graphene layers and graphene covered by sub-wavelength gratings, shown in Fig.~\ref{Fig0}.

The conductivity $\sigma(q,{\bf \omega})$ of drifting Dirac electron fluid is obtained by solving the kinetic equation for distribution function $f_{\bf p} = f^{(0)}_{\bf p} +\delta f_{\bf p}$ within linear response to the external electric field $\delta {\rm E}_{\bf q} = -i {\bf q} \delta \varphi_{\bf q}e^{i({\bf{qr}} - \omega t)}$:
\begin{equation}
    -i \omega \delta f_{\bf p} + i {\bf q v_p} \delta f_{\bf p} + i {\bf q} \delta \varphi_{\bf q} \partial_{\bf p} f^{(0)}_{\bf p} = \mathcal{C}_{ee}\{\delta f_{\bf p}\}, 
\end{equation}
here ${\bf v_p} = \partial_{\bf p}\epsilon_{\bf p}$ is the quasiparticle velocity. The carrier drift is encoded in zero-order distribution function, which we take in the local-equilibrium (hydrodynamic) form with velocity ${\bf u}_0$, Fermi energy $\varepsilon_F$ and temperature $T$, $f_0 = [1+e^{(\epsilon_p - {\bf p u}_0 - \varepsilon_F)/T}]^{-1}$. We restrict ourselves to collinear propagation of waves and carrier drift. 

The crucial step of the solution is the approximation of e-e collision integral $\mathcal{C}_{ee}\{\delta f_{\bf p}\}$ which does not enable any analytical treatment in its original form. We adopt $\mathcal{C}_{ee}\{\delta f_{\bf p}\}$ that pulls all perturbations of distribution function toward a local equilibrium $\delta f_{\rm hd}$ (but not to zero!~\cite{BGK_model,Ashcroft_conserving}) with a characteristic rate $\gamma_{ee} = \tau_{ee}^{-1}$:
\begin{gather}
    \mathcal{C}_{ee}\{\delta f_{\bf p}\} = -\gamma_{ee}(\delta f_{\rm p} - \delta f_{\rm hd}),\\
    \delta f_{\rm hd} = \delta\mu \partial_{\mu}f^{(0)}_{\bf p} + \delta {\bf u} \partial_{\bf u}f^{(0)}_{\bf p} + \delta T \partial_{T} f^{(0)}_{\bf p}.
\end{gather}
The main properties of true e-e scattering are encoded into the model, as the distribution modes corresponding to shift of particle number, momentum, and energy are not relaxed. The weights of these modes $\delta \mu$, $\delta {\bf u}$, and $\delta T$ are obtained from respective conservation laws for e-e collisions. These requirements lead us to a linear system of generalized HD equations which can be written symbolically as ${\hat M}{\delta \bf x} = {\delta \bf F}$. The vector $\delta {\bf x}$ contains unknown hydrodynamic parameters that can be arbitrary linear combinations $\delta \mu$, $\delta {u}$, and $\delta T$; $\delta \bf F$ is the vector of generalized force densities, and $\hat M$ is the dynamic matrix. The simplest form is achieved when relative perturbations of particle density $\delta n/n_0$, 'relativistic' velocity $\delta \beta = \delta u/v_0$ and mass density $\delta \rho/\rho_0$ are treated as unknowns. In this representation, the HD matrix and force vector take the form (Supporting Section II)
\begin{gather}
\label{HD-matrix}
 \hat{M}=\left( \begin{matrix}
   1-i{{\tilde{\gamma }}_{ee}}{I_{02}} & -i{{\tilde{\gamma}}_{ee}}{{\partial }_{\beta }}{I_{02}} & 0  \\
   0 & 1-\frac{2}{3}i{{\tilde{\gamma}}_{ee}}{\partial_\beta}{I_{13}} & {{\beta }_0}-\frac{2}{3}i{{\tilde{\gamma}}_{ee}}{I_{13}}  \\
   0 & {\beta}-i{\tilde{\gamma}_{ee}}{{\partial }_{\beta}}{I_{03}} & 1+\frac{\beta^2}{2}-i{\tilde{\gamma}_{ee}}{I_{03}}  \\
\end{matrix} \right) \\     
\label{HD_forces}
  {\delta {\bf F}}=-2\frac{e\delta \varphi }{m v_{0}^{2}}
      \left( 
  \begin{matrix}
   I_{12}-{\beta_0}{I_{02}}  \\
   I_{23}-{\beta_0}{I_{13}}  \\
  \frac{3}{2}\left({I}_{13}-{\beta_0}{I_{03}}\right)  \\
\end{matrix}
 \right)  
\end{gather}
where we have introduced 'relativistic mass' $m\approx \varepsilon_F/v_0^2$, the inverse Knudsen number $\tilde\gamma_{\rm ee} = (qv_0 \tau_{\rm ee})^{-1}$, and dimensionless functions $I_{nm}(a,\beta)$ of scaled frequency $a=(\omega+i\gamma_{ee})/qv_0$ and drift velocity $\beta$:
\begin{equation}
\label{Inm}
I_{nm}\left( a,\beta  \right)=\frac{{{\left( 1-\beta^2 \right)}^{m-\frac{1}{2}}}}{2\pi }\int\limits_{0}^{2\pi }{\frac{{{\cos }^{n}}\theta d\theta }{{{\left( 1-\beta \cos \theta  \right)}^{m}}\left( a-\cos \theta  \right)}}.   
\end{equation}

\begin{figure}[ht]
	\includegraphics[width=0.85\linewidth]{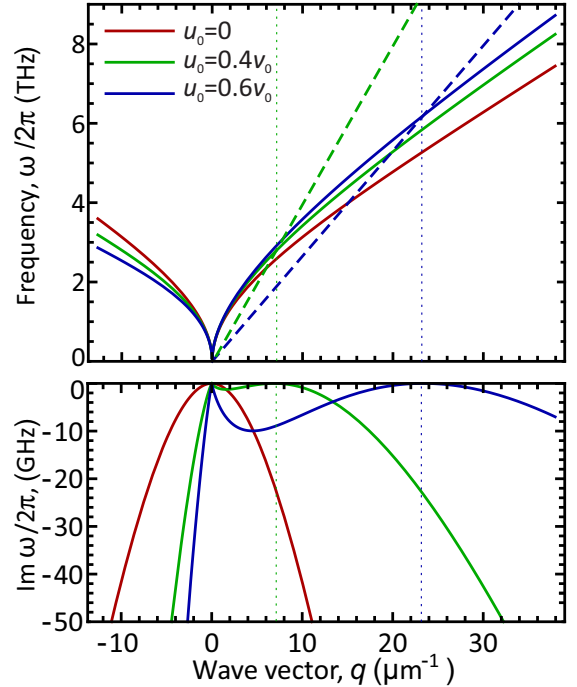}
	\caption{\label{Fig1} 
		Plasmon dispersion and damping in single graphene layer in the presence of electron drift and e-e collisions at different drift velocities. Positive wave vectors correspond to waves co-propagating with drift. The damping due to e-e collisions disappears at $q=0$ and $\omega u_0 = qv_0^2$ (shown with dashed lines). Fermi energy $\varepsilon_F = 25$ meV, background dielectric constant $\kappa = 5$}
\end{figure}

The system (\ref{HD-matrix}-\ref{HD_forces}) is the central result of this Letter. It provides an explicit expression for high-frequency non-local graphene conductivity $\sigma({\bf q},\omega)$ in the presence of carrier drift across the hydrodynamic-to-ballistic crossover (Supporting section II). It encloses numerous previous studies of graphene ac conductivity as limiting cases~\cite{Crossover,Kukhtaruk_2016,Briskot_NLHD}. Classical Navier-Stokes equations along with microscopic expression for viscosity are restored in the HD limit $\tilde{\gamma}_{ee}\gg 1 $ by expanding $I_{nm}$. In the opposite ballistic limit, $I_{nm}$ diverge at the boundary of single-particle excitations $\omega \rightarrow qv_0$. This divergence translates into singular conductivity at $\omega = qv_0$ and absence of plasmon modes below the singular line~\cite{Lundeberg_Nonlocal}.

{\it Drift-induced Doppler shift and plasmon undamping.} Several non-trivial plasmonic effects appear already in isolated graphene layer in the presence of drift due to the breakdown of Galilean invariance. The latter is readily seen from the generalized hydrodynamic system (\ref{HD-matrix}-\ref{HD_forces}) as the wave frequency $\omega$ and drift velocity $u_0$ appear therein not only in combination $\omega - q u_0$, as it should be for massive electrons. 

The first such effect is anomalous Doppler splitting betwеen frequencies of up- and downstream plasmons $\Delta\omega^\pm$. It is always below the conventional value of $2 q u_0$; in the hydrodynamic limit it is exactly one half of it. In the ballistic limit
\begin{equation}
\label{Ballistic-shift}
\Delta\omega_{\rm bal}^\pm = 2 q u_0 (s^2-1)(\sqrt{s^2-1}-s)^2, 
\end{equation}
where $s = \omega/qv_0$ is the ratio of wave phase velocity and Fermi velocity. The ballistic Doppler shift approaches zero as the wave velocity approaches $v_0$; it stems from singular ballistic conductivity at $\omega = q v_0$.

Much more surprising is the wave damping due to e-e collisions, which is shown on the bottom panel of Fig.~\ref{Fig1}. The damping of upstream wave continuously increases with the drift speed. The damping of downstream wave for finite $u_0$ approaches zero at some peculiar frequencies satisfying $\omega u_0 = q v_0^2$, and then continues to grow. Far away from 'undamping point', the imaginary part of frequency is proportional to $q^2$, as it should be for the viscous damping.

The origin of undamping points can be traced back to the excitation of distribution modes $\delta f$ that are insensitive to e-e collisions. This is readily seen in 'boosted coordinates' $(\tilde p,\theta)$, where $\tilde p = p(1-\beta \cos\theta)$~\cite{Kashuba_NLHD}. In the absence of collisions and at $T/\varepsilon_F \ll 1$, electric field excites the distributions
\begin{equation}
    \delta f\propto \frac{\cos\theta \delta( \tilde p v_0 - \varepsilon_F)}{[1-(qv_0/\omega)\cos\theta][1-\beta\cos\theta]^2},
\end{equation}
which do not generally coincide with zero modes of $\mathcal{C}_{ee}$ and are therefore relaxed. But at special points $qv_0/\omega = \beta$, the excited distribution coincides with the hydrodynamic momentum mode of $\mathcal{C}_{ee}$:
\begin{equation}
    \delta f\propto \frac{\cos\theta \delta( \tilde p v_0 - \varepsilon_F)}{[1-\beta\cos\theta]^3},
\end{equation}
It implies that collisions do not have any effect on these modes, and relaxation is absent. We note here that undamping occurs not only for plasmons, but the whole conductivity becomes dissipationless ($\sigma'(q,\omega) = 0$) at these special frequencies and wave vectors.

\begin{figure}[ht]
	\includegraphics[width=0.85\linewidth]{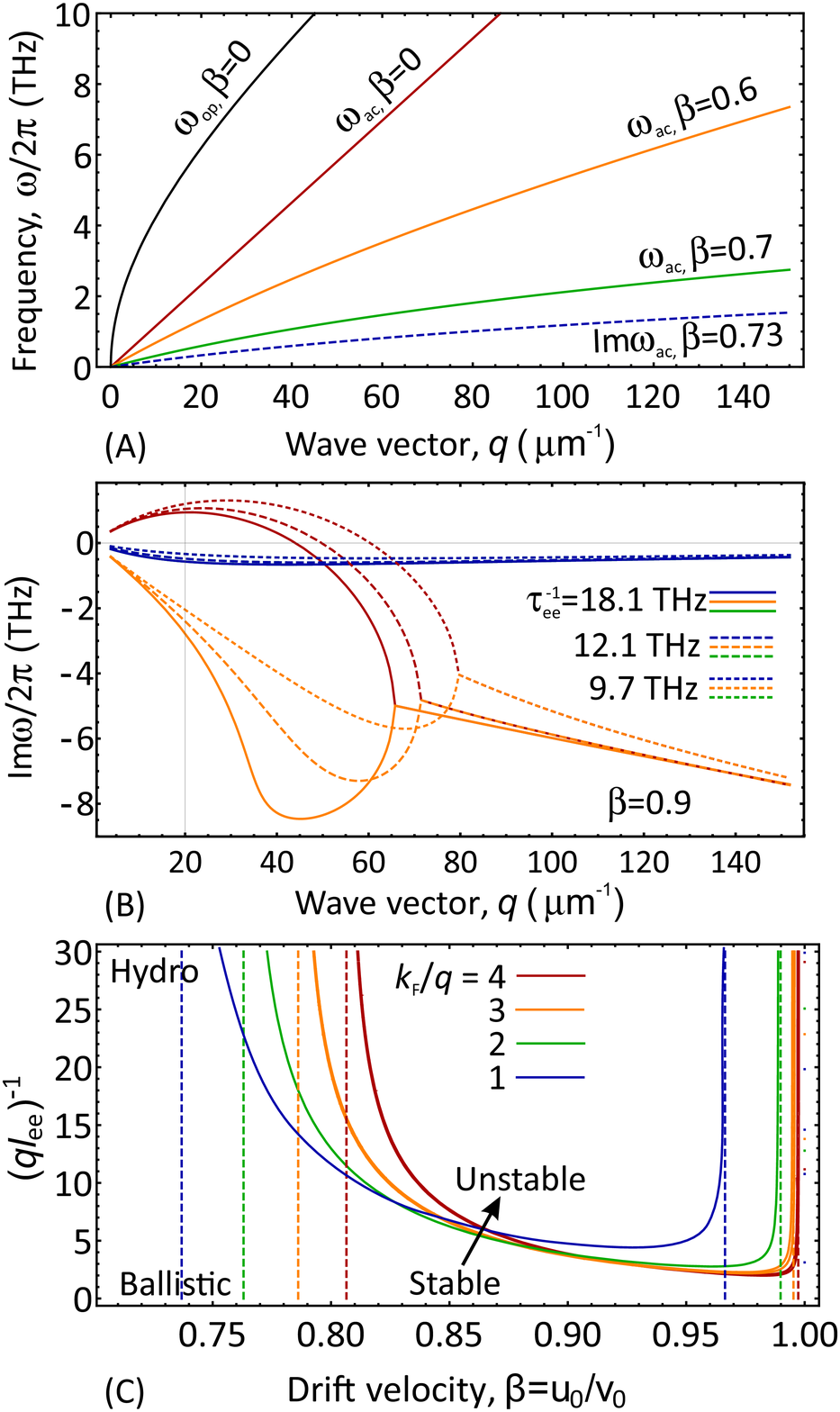}
	\caption{\label{Fig2} 
		Plasmons in graphene double layers with counter-streaming flows
		(A) Evolution of spectra in HD regime with increasing drift velocity $\beta$ leading to the shift of acoustic mode frequency down to zero and subsequent instability (B) Damping/growth rates of plasmons at different values of e-e scattering rate, demonstrating suppression of instability with reduced scattering (C) Stability diagram of the double-layer setup: critical value of velocity and inverse Knudsen number at which plasma waves become unstable. In panels (A) and (B), interlayer distance $d=1$ nm, Fermi energy $\varepsilon_F = 25$ meV. In panel (C), interlayer coupling $e^{-2qd}$ is set to 0.86
		}
\end{figure}

{\it Instabilities of counter-streaming flows.} Cerenkov-type plasmon instability is most simply achieved in a double layer setup where electron velocities in two layers point in opposite direction. The theory of such instabilities developed for massive electrons~\cite{krasheninnikov1980instabilities} was recently erroneously applied to graphene in the ballistic regime~\cite{Neagtive_LD,Gumbs_Instability}. Here, we find that the very presence of such instabilities depends strongly on transport regime in the two layers.

 When both layers are ballistic, $\omega\tau_{ee} \gg 1$, instabilities are absent despite negative real part of conductivity in the Cerenkov domain, $\omega \lesssim q u_0$~\cite{Comment}. The symmetric (optical) plasmon mode is unaffected by drift, while the frequency of asymmetric one is pulled by current toward lower frequencies. However, it cannot be decreased below the boundary of single-particle excitations $\omega = qv_0$ due to singular non-local response of graphene, as given by Eq.~\ref{Ballistic-shift}. As a result, the gain region encloses no plasmon eigenmode.

The acoustic mode, however, readily reaches the region of Cerenkov gain in the hydrodynamic regime, as shown in Fig.~\ref{Fig2} A. The non-local dielectric response of graphene in this regime is no more singular, and the wave frequency unimpededly passes through $\omega = qv_0$ border. Above the critical velocity, the mode aperiodically growing, i.e. ${\rm Im}\omega > 0$, ${\rm Re}\omega = 0$. The range of velocities for observation of instabilities in HD regime is located between $\beta^-_{\rm th}$ and $\beta^+_{\rm th}$, 
\begin{equation}
\label{Crit-velocity}
    \beta^\pm_{\rm th} = \frac{v_0}{\sqrt{2}}\sqrt{\frac{q^2v_0^2 + 2\omega^2_p(1 \pm e^{-qd})}{q^2v_0^2 + \omega^2_p(1 \pm e^{-qd})}},
\end{equation}
where $\omega_p = (2 \pi n e^2|q|/m)^{1/2}$ is the plasma frequency in an isolated graphene layer. $\beta^-_{\rm th}$ has a natural lower bound $v_0/\sqrt{2}$ that coincides with sound velocity. Interestingly, the threshold velocity weakly depends on carrier density as far as layers are closely bound ($qd \ll 1$).

The growth rate of unstable modes is going down as the e-e collision frequency is reduced, shown in Fig.~\ref{Fig2} B. In the weak HD regime, $ql_{\rm ee} \lesssim 1$, the reduced growth rate can be attributed to increased viscous damping. However, outside of the hydrodynamic domain $ql_{\rm ee} \gtrsim 1$, the instabilities do not re-appear as the velocity of acoustic modes is forced to lie above the velocity of carrier drift.

The full stability diagram of counter-streaming double-layer system is calculated in Fig.\ref{Fig2} C: the values of drift velocity and e-e collision frequency above the threshold lines correspond to unstable modes. Remarkably, the e-e collision frequency in the dispersion relation appears scaled to $q v_0$. It implies that the only parameter governing the transition between HD and ballistic regimes is the Knudsen number $qv_0/\gamma_{\rm ee} = ql_{\rm ee}$, where $l_{\rm ee}$ is electron free path with respect to e-e collisions. As a result, instabilities can always be observed in clean systems of sufficiently large length.


{\it Distributed-feedback plasmon lasing.} A highly resonant instability leading to electromagnetic emission can be observed in graphene with conducting substrate covered by a metal grating. Such setup is commonly used for spectroscopy of plasmon resonance in 2d electron systems~\cite{Allen_Plasmons}. The reflectance spectrum of such setup is calculated using the formalism of \cite{Mikhailov_Instability,Matov_Emission} with graphene conductivity found from Eqs.~(\ref{HD-matrix}-\ref{HD_forces}) as a building block.

\begin{figure}[ht]
	\includegraphics[width=0.85\linewidth]{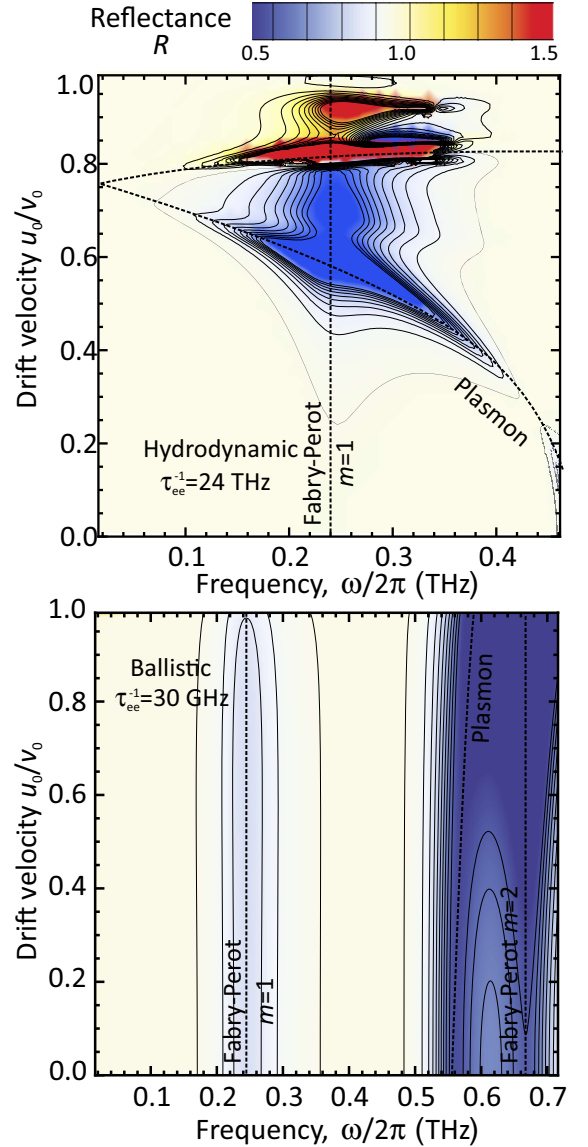}
	\caption{\label{Fig3} 
		Evolution of reflectance spectra of graphene covered by plasmonic grating with increasing carrier drift velocity in the hydrodynamic (A) and ballistic (B) regimes. Bright regions correspond to the excitation of plasmon and Fabry-Perot modes. At large drift velocity, plasmon-enhanced absorption ($R<1$) turns to plasmon-enhanced amplification ($R>1$). The reflection coefficient further diverges at the crossing of Fabry-Perot and plasmon resonance. Parameters: $\varepsilon_F = 50$ meV, $d=3$ nm, $D = 100$ $\mu$m, $\kappa = 12$, grating period is 2 $\mu$m, filling factor is $1/2$}
\end{figure}

The electromagnetic response of grating-coupled graphene differs for hydrodynamic and ballistic regimes already in the absence of dc current. Namely, the frequencies of plasmonic dips are reduced by e-e collisions (Fig. 4). When passing direct current in 2DES, the absorption peak is split by Doppler effect~\cite{Matov_Emission,Mikhailov_Instability,Wenger_CurrentControlled}. With increasing current, the distinctions between HD and ballistic regimes become more drastic.

The Doppler shift in the ballistic regime is so weak and the singular non-local response at $\omega = qv_0$ is so strong that the plasmon frequency is almost unaffected by current (shown in Fig.~\ref{Fig3} B). On the contrary, the plasmon frequency in the hydrodynamic regime passes to zero frequency unimpededly (Fig.~\ref{Fig3} A). At higher current, the resonant frequency grows again, but the reflection coefficient exceeds unity. The negative absorption is associated with generation of evanescent waves by the grating that fall into the negative conductivity domain of 2DES; the effect is enhanced if frequency corresponds to plasmon resonance.

Both absorption and amplification effects are enhanced if the distance $D$ between 2DES and conducting substrate satisfies the anti-reflection condition, $D = \lambda_0/4\sqrt{\kappa}$, where $\kappa$ is the background dielectric constant. The enhancement occurs once the eigen-frequency of Fabry-Perot cavity formed in the vertical direction coincides with the frequency of plasmon. The interaction of plasmon at high current with Fabry-Perot mode leads to divergent reflection coefficient. The divergence implies that such a mode can grow without external stimulus until it is stabilized by nonlinear effects, as it occurs in the distributed feedback lasers.

{\it Discussion and conclusions.} 
We now argue that Cerenkov-type plasmon instability in Dirac materials in HD regime and its absence in ballistic regime are linked to singular structure of conductivity $\sigma(q,\omega)$ and independent of particular dielectric environment. These considerations should be applicable both to 2D and 3D Dirac materials, where conductivity has square-root and log-singularities~\cite{lv2013dielectric}, respectively, at $\omega = q v_0$. 

The TM plasmon modes of arbitrary structure exist in the domain of positive imaginary part of conductivity ${\rm Im}\sigma(q,\omega)>0$ which, in the ballistic regime, lies above the singularity, $\omega > q v_0$. The Cerenkov domain $\omega < qu_0$ lies below the singularity. The position of singularity is insensitive to carrier drift. Therefore, one cannot thread the plasmon modes through the singularity by a continuous change of parameter $u_0$. This, however, becomes possible in the hydrodynamic regime, where the singularity is removed from real frequency axis due strong to e-e collisions. These arguments do not apply to combinations of Dirac and parabolic-band materials (e.g. graphene parallel to bulk collisionless plasma), where joint plasmon modes can exist at $\omega < qu_0$. 


The predicted effects can be readily tested experimentally. The anomalous Doppler shifts of plasmons in graphene can be measured with Raman spectroscopy, as it was done for III-V based 2DES~\cite{Raman_plasmons}. The plasmon instabilities, both in double layer~\cite{Gribnikov_2stream} and grating-gate setup can result in oscillatory current regimes and emission of terahertz radiation. Such emission can be distinguished from hot-plasmonic emission~\cite{Tsui_emission} by the presence of well-defined threshold current~\cite{ElFatimy_THz_emission}. Reflectance spectroscopy of grating-gated 2DES is another convenient tool to study Doppler shift and wave amplification~\cite{Otj_grating}.


{\it Acknowledgement.} This work was supported by the grants 18-37-20058/18 and 16-29-03402/18 of the Russian Foundation for Basic Research. The author thanks Denis Fateev for helpful discussions and Mikhail K. Maslov for assistance at the early stage of the work.

\bibliography{apssamp}

\begin{widetext}
\section{Supporting information}
\subsection{Hydrodynamic transport regime}
The hydrodynamic equations for electron drift velocity $\bf u$, chemical potential $\mu$ and temperature $T$ can be obtained by supplying the local-equilibrium distribution function
\begin{equation}
    f_0 = [1+e^{(\epsilon_p - {\bf p u}_0 - \mu)/T}]^{-1}
\end{equation}
into kinetic equation and integrating it by phase space multiplied by $1$, $\bf p$ and $\epsilon_{\bf p}$. Though a more accurate derivation with explicit account of collisions is possible (see section II), such simplistic derivation may be convenient for analysis of nonlinear effects. The resulting equations can be presented as
\begin{gather}
    \partial_t n + \partial_{\bf r}(n{\bf u})=0,\\
    \partial_t(\rho u_i) + \partial_{x_j}\Pi_{ij} = e n \partial_{x_i}\varphi,\\
    \partial_t\varepsilon + \partial_{\bf r}(\rho v_0^2 {\bf u}) - e n {\bf u}{\partial_{\bf r}\varphi} = 0. 
\end{gather}

Above, $n$ is the density of electrons, $\rho$ is the equivalent of mass density, $\varepsilon$ is the internal energy density, and $\Pi_{ij}$ is the stress tensor. At given value of chemical potential $\mu$, all these quantities depend on drift velocity $\beta = u / v_0$. At the same time, they can be expressed via their values in the absence of drift, i.e. at $\beta = 0$:
\begin{gather}
\label{Eqs-of-state}
    n = \frac{n_{\beta = 0}}{[1-\beta^2]^{3/2}},\qquad     \rho = \frac{\rho_{\beta = 0}}{[1-\beta^2]^{5/2}},\\
    \varepsilon = \varepsilon_{\beta = 0} \frac{1 + \beta^2/2}{[1-\beta^2]^{5/2}},\\ \rho_{\beta = 0}v_0^2 = \frac{3}{2}\varepsilon_{\beta = 0}
    \Pi_{xx} = \frac{\varepsilon_{\beta = 0}}{2}\frac{1+2\beta^2}{[1-\beta^2]^{5/2}},\qquad \Pi_{yy} = \frac{\varepsilon_{\beta = 0}}{2}\frac{1 - \beta^2}{[1-\beta^2]^{5/2}}
\end{gather}

To avoid dealing with 'relativistic factors' of the type $[1-\beta^2]^{\alpha}$, it is convenient to consider density $n$, mass density $\rho$ and velocity $\beta$ as hydrodynamic variables. In these variables, the Euler and heat balance equations take on the closed form (assuming one-dimensional motion along $x$-axis)
\begin{gather}
    \partial_t(\rho u_x) + \frac{1}{3}\partial_{x}[\rho  (v_0^2 + 2 u^2)] = e n \partial_{x}\varphi,\\
   \frac{2}{3} \partial_t(\rho  [v_0^2 + u^2/2]) + \partial_{\bf r}(\rho v_0^2 {\bf u}) - e n {\bf u}{\partial_{\bf r}\varphi} = 0. 
\end{gather}
The equations can now be linearized to find the conductivity of drifting Dirac electrons in the hydrodynamic regime, $n = n_0 + \delta ne^{i(kx-\omega t)}$, $u = u_0 + \delta u e^{i(kx-\omega t)}$, $\rho = \rho_0 + \delta \rho e^{i(kx-\omega t)}$. This results in the system of equations $\hat M_{\rm hd}\delta{\bf x} = \delta {\bf F}_{\rm hd}$, with the matrix and right-hand side given by
\begin{gather}
\label{HD-matrix-2}
 \hat{M}_{\rm hd}=\left( \begin{matrix}
   -i(\omega - q u_0) & i q & 0  \\
   0 & -i (\omega - \frac{4}{3}qu_0) & - i (\beta_0\omega - \frac{q}{3})  \\
   0 & - i (\beta_0 \omega - \frac{3}{2}q) & -i(\omega - \frac{3}{2}qu_0)  \\
\end{matrix} \right) \\     
\label{HD_forces-2}
  {\delta {\bf F}}_{\rm hd}=\frac{e\delta \varphi }{m v_{0}^{2}}
      \left( 
  \begin{matrix}
   0 \\
   i q v_0  \\
  \frac{3}{2}i q u_0  \\
\end{matrix}
 \right)  
\end{gather}
the hydrodynamic mass $m$ si given by
\begin{equation}
    m = \frac{\rho_0}{n_0} \approx \frac{\varepsilon_F}{v_0^2 - u_0^2}.
\end{equation}

As a final step, it is possible to solve the system (\ref{HD-matrix-2}-\ref{HD_forces-2}) for density $\delta n$ and velocity $\delta u$ to find the polarizability and conductivity
\begin{gather}
\label{HD_polarization}
    \Pi = -\frac{n q^2 (1-\beta_0^2)}{m \left[\omega^2(1-\frac{\beta_0^2}{2}) - \frac{q^2v_0^2}{2}(1-2\beta^2) - 2 q u_0 \omega\right]},\\
    \sigma = -\frac{ie^2\omega}{q^2}\Pi.
\end{gather}

At zero drift velocity, the polarzation of graphene acquires a simple form common for bulk solids:
\begin{equation}
    \Pi = -\frac{n q^2 /m}{\omega^2 - v_s^2 q^2},
\end{equation}
where $v_s = v_0/\sqrt{2}$ is the sound velocity.

\subsection{Polarizability and conductivity for drifting electrons at the hydrodynamic-to-ballistic crossover}
The solution of kinetic equation with electron-electron collisions reads
\begin{equation}
\label{Distribution_funtion}
    \delta f = \frac{i\nu_{ee}\delta f_{\rm hd} - i e\delta\varphi {\bf q} \partial_{\bf p}f_0}{\omega+i\nu_{ee}-{\bf qv_p}},
\end{equation}
where the perturbed hydrodynamic distribution reads
\begin{equation}
    \delta f_{hd} = \delta\mu \partial_{\mu}f_0 + \delta {\bf u} \partial_{\bf u}f_0 + \delta T \partial_{T}f_0,
\end{equation}
and the unperturbed drifting distribution is 
\begin{equation}
    f_0 = \left[1+\exp\left\{ \frac{\epsilon_p - {\bf p u}_0 - \mu}{T}\right\}\right]^{-1}.
\end{equation}
The local-equilibrium Fermi energy $\mu$, drift velocity ${\bf u}_0$ and temperature $T$ should be determined from the solution of dc transport equations from the known drain and gate voltages. We leave this solution for further work. The variations of local-equilibrium parameters due to ac field $\delta \mu$, $\delta {\bf u}$ and $\delta T$ should be obtained by requiring particle, momentum and energy conservation upon e-e collisions:
\begin{equation}
\label{Conservation_laws}
    \sum_{\bf p}{[\delta f- \delta f_{\rm hd}]} = 0, \qquad     
    \sum_{\bf p}{{\bf p}[\delta f- \delta f_{\rm hd}]} = 0, \qquad
        \sum_{\bf p}{\epsilon_{\bf p}[\delta f- \delta f_{\rm hd}]} = 0.
\end{equation}
Substituting the distribution function (\ref{Distribution_funtion}) into conservation laws (\ref{Conservation_laws}), we can obtain a set of generalized hydrodynamic equations for determination of Fermi energy, drift velocity and temperature. In the course of evaluation, one encounters the following integrals
\begin{equation}
\sum_{\bf p}{\frac{p^{m-1} \cos^n\theta_{\bf p} f_0({\bf p})}{a-\cos\theta_{\bf p}} },\qquad a=\frac{\omega+i\nu_{ee}}{qv_0}.    
\end{equation}
Anisotropy of $f_0({\bf p})$ introduces complications that can be handled via the change of momentum
\begin{equation}
    \tilde p = p(1 - u_0/v_0 \cos\theta_{\bf p}),
\end{equation}
after which the integrals over momentum modulus and angle are decoupled
\begin{equation}
\sum_{\bf p}{\frac{p^{m-1} \cos^n\theta_{\bf p} f_0({\bf p})}{a-\cos\theta_{\bf p}} } = \frac{1}{(2\pi)^2}\int_0^{\infty}{d\tilde p \tilde p^m f_F(\tilde p)} \int_0^{2\pi}{\frac{\cos^n\theta d\theta}{(a-\cos\theta)(1-\beta \cos\theta)^m}}.    
\end{equation}
Above $f_F$ is already an isotropic Fermi function with the same values of $\mu$, $T$ and zero drift velocity. We introduce the notation
\begin{equation}
    J_{nm} = \frac{1}{2\pi}\int\limits_0^{2\pi}{\frac{\cos^n\theta d\theta}{(a-\cos\theta)(1-\beta \cos\theta)^m}}.
\end{equation}
The integral is evaluated by setting $z=e^{i\theta}$ and computing the residues in the poles inside the unit circle $|z|=1$. The remaining integrals over momentum modulus can be expressed via electron density $n$ and energy density $\varepsilon$ at zero drift velocity $\beta\equiv u_0/v_0$. This results in generalized hydrodynamic system:
\begin{gather}
    \delta n = \frac{i\nu_{ee}}{qv_0}\delta[n_{\beta = 0} J_{02}] + i \frac{e \delta\varphi}{qv_0} \frac{\partial n_{\beta = 0}}{\partial \mu} [qv_0  J_{12} - {\bf qu}_0 J_{02} ],\\
    \delta{{\bf P}v_0} = \frac{i \nu_{ee}}{qv_0} \delta[\varepsilon_{\beta = 0} J_{13}]+ i \frac{e \delta\varphi}{qv_0} 2n_{\beta = 0} [qv_0  J_{23} - {\bf qu}_0 J_{13} ],\\
    \delta{\varepsilon} = \frac{i \nu_{ee}}{qv_0} \delta[\varepsilon_{\beta = 0} J_{03}]+ i \frac{e \delta\varphi}{qv_0} 2n_{\beta = 0} [qv_0  J_{13} - {\bf qu}_0 J_{03} ].
\end{gather}

Finally, to justigy the form (\ref{HD-matrix}-\ref{HD_forces}) of the main text, we express the quantities $n_{\beta=0}$ and $\varepsilon_{\beta=0}$ through $n$ and $\varepsilon$ at finite drift velocity using the equations of state (\ref{Eqs-of-state}). We also introduce another set of dimensionless integrals
\begin{equation}
    I_{nm} = (1-\beta^2)^{m-1/2}J_{nm},
\end{equation}
and inverse Knudsen number $\tilde \gamma_{ee} = (qv_0 \tau_{ee})^{-1}$. Generally, the right-hand side of generalized hydrodynamic equations has the form 
\begin{equation}
  {\delta {\bf F}}=-2e\delta \varphi 
      \left( 
  \begin{matrix}
   \frac{I_{12}-{\beta_0}{I_{02}}}{m_{k,\beta=0} v_{0}^{2}}  \\
   \frac{I_{23}-{\beta_0}{I_{13}}}{m_{hd,\beta=0} v_{0}^{2}}  \\
  \frac{3}{2}\frac{{I}_{13}-{\beta_0}{I_{03}}}{m_{hd,\beta=0} v_{0}^{2}}  \\
\end{matrix}
 \right)
\end{equation}
where the 'kinetic' and 'hydrodynamic' masses have the form
\begin{equation}
    m_{k,\beta=0} = \frac{n_{\beta=0}}{v_0^2 \partial n_{\beta=0}/\partial\mu},\qquad m_{hd,\beta=0} = \frac{\rho_{\beta=0}}{n_{\beta=0}}.
\end{equation}
In the main text, we neglect the difference between these masses which is justified in the degenerate limit $T/\epsilon_F \ll 1$. A small difference between these masses may result in extra plasmon damping due to relaxation of velocity modes by e-e collisions in non-parabolic bands~\cite{Crossover}.

The formal solution for polarizability is readily derived from generalized hydrodynamic system~(\ref{HD-matrix}-\ref{HD_forces}). Denoting the elements of hydrodynamic matrix as $M_{ij}$ and components of generalized force vector as $\delta F_i$, we find
\begin{equation}
    \Pi (q,\omega) = \frac{1}{e\delta\varphi}\frac{\delta F_1}{M_{11}} +\frac{M_{12}}{M{11}}\frac{M_{33}\delta F_2 - M_{23}\delta F_3}{M_{23}M_{32} - M_{22}M_{33}}
\end{equation}

\subsection{Evaluation of auxiliary spatially dispersive integrals}
Upon obtaining the generalized hydrodynamic equations, on encounters the following integrals
\begin{equation}
    J_{nm} = \frac{1}{2\pi}\int\limits_0^{2\pi}{\frac{\cos^n\theta d\theta}{(a-\cos\theta)(1-\beta \cos\theta)^m}}.
\end{equation}
At any given integers $n$ and $m$, they are evaluated by passing to the complex variable $z=e^{i\theta}$. This results in
\begin{equation}
    J_{nm} = \frac{1}{2\pi i}\int\limits_{|z|=1}{\frac{(z+z^{-1})^n dz}{2^n z [a - (z+z^{-1})/2][ 1-\beta (z+z^{-1})/2]^m}}.
\end{equation}
The integrand has poles at the points:
\begin{equation}
    z_0=0,\qquad z^{(a)}_{\pm} = a \pm \sqrt{a^2-1}, \qquad z^{(\beta)}_{\pm} = \frac{1\pm\sqrt{1-\beta^2}}{\beta}.
\end{equation}
Among these points, $z^{(\beta)}_{-}$ and $z_0$ lie inside the unit circle $z=1$, $z^{(a)}_{-}$ lies inside the unit circle for $\rm{Re} a > 0$ and $z^{(a)}_{+}$ -- for $\rm{Re} a < 0$. These statements are independent of sign of imaginary part of $a$.

Evaluation is completed by computation of integrand residues at these poles. As a result, we arrive at the following expressions:
\begin{gather}
\label{Integrals}
    J_{03} = \frac{1}{2 (a \beta -1)^3} \left\{\beta \frac{  \left(a^2+2\right) \beta ^4+\left(2 a^2-5\right) \beta ^2-6 a \beta +6 }{\left(1-\beta
   ^2\right)^{5/2}}+ 2 i \frac{\text{sign} \rm{Im} a }{\sqrt{1-a^2}}\right\},\\
   J_{02} = \frac{1}{(a
   \beta -1)^2} \left\{\frac{\text{sign} \rm{Re} a}{\sqrt{a^2-1}}+ \beta \frac{ \beta  (a+\beta )-2 }{\left(1-\beta ^2\right)^{3/2}}\right\},\\
   J_{12} = \frac{1}{{(a \beta -1)^2}} \left\{a \frac{\text{sign} \rm {Re} a}{\sqrt{a^2-1}}-\frac{1-a \beta ^3}{\left(1-\beta ^2\right)^{3/2}}\right\}.
\end{gather}
The remaining necessary integrals can be obtained with recurrence relations
\begin{equation}
    \frac{\partial J_{nm}}{\partial \beta} = m J_{n+1,m+1}.
\end{equation}
As apparent from the forms (\ref{Integrals}), there is a singularity at $a = \rightarrow 1$. As the parameter $a$ has finite imaginary part ($a = (\omega + i\gamma_{ee})/qv_0$), the singularity is present only in the ballistic regime (i.e. at $\omega \gg \gamma_{ee}$). Finite strength of e-e collisions softens the singularity. 

The integrals also diverge at $\beta \rightarrow 1$, a situation close to that in special relativity. However, this divergence can be re-absorbed into definitions of particle and mass density (\ref{Eqs-of-state}), so that the resulting generalized hydrodynamic equations are free of divergences.

There is a spurious singularity at $a \beta \rightarrow 1$, however, a closer inspection reveals that it is compensated by the zero value of the numerator. An only special property of the computed dielectric response at $a \beta \rightarrow 1$ is the absence of dissipation, as discussed in the main text.

\subsection{Analysis of instabilities in the double-layer system}
The dispersion law for plasmons in a double-layer structure separated by distance $d$ reads
\begin{equation}
\label{Double-layer}
   \epsilon_{2l}(q,\omega,\beta)\equiv (1 + V_0 \Pi_+) (1 + V_0 \Pi_-) - V^2_0 \Pi_- \Pi_+ e^{-2|q|d} = 0.
\end{equation}
Here $\Pi_+$ and $\Pi_-$ are the polarizabilities of individual top and bottom layers, $V_0 = 2\pi e^2/\kappa |q|$ is the Fourier transform of Coulomb interaction and $\kappa$ is the background dielectric constant. Substitution of hydrodynamic polarizability (\ref{HD_polarization}) results in biquadratic equation with two eigenmodes
\begin{equation}
\label{HD-modes-double-layer}
    \frac{\omega^2_{\pm}}{q^2v_0^2} = \frac{  2 \left(\beta ^4-3 \beta ^2+2\right)  s_p^2 + 2 \beta ^4-3 \beta ^2+2  \pm 2 \sqrt{2 \beta ^2 \left(\beta
   ^2-1\right)  \left(\left(\beta ^2-1\right) + \left(\beta ^2-2\right) s_p^2 \right)+\left(\beta ^4-3 \beta ^2+2\right)^2 s_p^4 e^{-2 d
   q}}}{\left(\beta ^2-2\right)^2},
\end{equation}
here we have introduced the dimesionless 'plasmon phase velocity' 
\begin{equation}
    s_p = \frac{\omega_p}{qv_0}, \qquad \omega_p = \sqrt{\frac{2\pi n e^2 |q|}{\kappa m_{hd}}}.
\end{equation}
The signs $+$ and $-$ in the absove dispersion can be traced back to optical and acoustic modes of the double-layer structure in the absence of drift, respectively. Indeed, at $\beta = 0$ one obtains
\begin{equation}
    \omega^2_{\pm} = \frac{q^2v_0^2}{2} + \omega_p^2(1 \pm e^{-2qd}).
\end{equation}

The instability emerges as the frequency of acoustic mode in (\ref{HD-modes-double-layer}) passes through zero with increasing the drift velocity. This occurs at
\begin{equation}
\beta^\pm_{\rm th} = \frac{v_0}{\sqrt{2}}\sqrt{\frac{q^2v_0^2 + 2\omega^2_p(1 \pm e^{-qd})}{q^2v_0^2 + \omega^2_p(1 \pm e^{-qd})}}. 
\end{equation}

The solution of dispersion relation (\ref{Double-layer}) with polarizability including e-e collisions shows that instability sets on once the acoustic mode frequency crosses zero, i.e. at $\rm Re \omega_-  = 0$. A direct verification of this fact is challenging, but the experience of numerical solutions tells that it is the case. The stability diagram in Fig.~\ref{Fig2} (C) was therefore obtained by numerical solution of $\epsilon_{2l}(q,0,\beta)=0$.

We note that the pattern of instabilities can be much richer if the carrier densities in the two layers are non-equal. In this case, the instability does not necessarily set on once the mode frequency crosses zero. A detailed analysis of these cases will be presented elsewhere.

\subsection{Diffraction on grating-gated graphene}
We consider the diffraction of an electromagnetic wave normally incident on graphene covered by a metal grating. The electric field is polarized along the $x$-axis, i.e. perpendicular to the gratings. The metal grating is assumed to be infinitely thin, its surface conductivity $\sigma_{\rm m}$ exceeds the velocity of light, $\sigma_{\rm m} \gg c$, and is set to infinity in the following calculation. The grating-to-graphene distance $d$ is well below the grating period $a$ and with $W = f a$, where $f$ is the filling factor. This ensures efficient coupling of evanescent waves generated by the grating to the surface plasmons. We also assume that the structure is globally gated with a highly conducting substrate, the distance to back gate $z_0$ is the largest length scale in the problem, $z_0 \gg W \gg d$. We tune $z_0$ to accommodate nearly quarter of wavelength in the substrate material, $z_0 \sim \lambda_0/4n_{\rm sub}$, where $\lambda_0$ is the free-space wavelength.

Due to uniformity of 2DES is the $x$-direction, the diffraction problem can be formulated on surface current in the grating $j_s(x)$, $x \in [0;W]$, while all the information about 2DES is accommodated in the Green's function of electromagnetic problem. This results in the following integral equation
\begin{equation}
\label{Diffraction-problem}
    \frac{j_s(x)}{\sigma_{\rm m}} = \mathcal{E}_0 + \int\limits_{0}^{W}{dx' Z(x-x') j_s(x')},
\end{equation}
where $\mathcal{E}_0$ is the field in the grating plane $z=0$ {\it in the absence of grating}. It can be presented as
\begin{equation}
\mathcal{E}_0 = \mathcal{E}_{\rm inc}(1 - r_{2d}),
\end{equation}
where $r_{2d}$ is the reflection coefficient of bottom-gated 2DES without grating, and $\mathcal{E}_{\rm inc}$ is the electric field in the incident wave. 

The impedance kernel $Z(x-x')$ is obtained as follows. First, one finds electric field induced at $z=0$ by $G$-th spatial Fourier harmonic of surface current passing in the grating plane:
\begin{equation}
\mathcal{E}_{{\rm ind}{G}} = j_{s,{G}}  Z_{G},
\end{equation}
the function $Z_{G}$ is easily obtained by plane-wave matching or transfer-matrix methods. Then $Z(x-x')$ is the inverse Fourier transform of $Z_{G}$ with the wave vectors running across the reciprocal wave vectors of the grating, $G_n = 2\pi n/a$:
\begin{equation}
\label{Impedance_kernel}
    Z(x-x') = \sum\limits_{n=-\infty}^{+\infty}{Z_{G_n} e^{-iG_n(x-x')}}.
\end{equation}

One should distinguish the cases $n=0$ (corresponding to the normally incident propagating wave) and $|n|\ge 1$ (corresponding to evanescent waves generated by grating):
\begin{gather}
\label{Impedance1}
   Z^{-1}_{|G|\ge 2\pi/a}  = \frac{i \omega }{4 \pi  |G| }\left\{ 1 - \varepsilon_{\rm sub} + \frac{2 \varepsilon_{\rm sub}  \left(1+\frac{2 i \pi  \sigma_{2d} |G| }{\omega \varepsilon_{\rm sub} }\right)}{1+\frac{2 i \pi  \sigma_{2d} |G| 
   \left(1-e^{-2 d |G| }\right)}{\omega \varepsilon_{\rm sub} }}  \right\},\\
\label{Impedance2}
   Z^{-1}_{G=0} = \frac{\omega}{4 \pi  k_1}  \left\{\frac{2 \varepsilon_{\rm sub}   \left(1-\frac{2 \pi  k_1 \sigma}{\omega  \varepsilon_{\rm sub}  }  \left(-1+e^{2 i  k_1 (z_0 - d)}\right) \right)}{\frac{2 \pi  k_1 \sigma}{\omega  \varepsilon_{\rm sub}  }
    \left(\left(-1+e^{-2 i d k_1}\right) e^{2 i k_1 z_0 }+e^{2 i d
   k_1}-1\right)+e^{2 i k_1 z_0}-1}-\frac{k_1}{k}+\varepsilon_{\rm sub} \right\};
\end{gather}
here $k=\omega/c$ and $k_1 = \omega \sqrt{\varepsilon_{\rm sub}}/c$ are the wave vectors in vacuum and the substrate material.

It is important to note that impedance kernel (\ref{Impedance_kernel}) with Fourier components (\ref{Impedance1}-\ref{Impedance2}) diverges at large $G$. This is associated with singularities of electric field near the keen edges of thin metal stripe carrying uniform current. The divergence can be cured in two ways. As a first possibility, one can transform Eq.~\ref{Diffraction-problem} into a second-order differential with respect to $x$ and impose zero boundary conditions on current at the edges \begin{equation}
    j_s(x=0) = j_s (x=W) = 0.
\end{equation}
The Fourier components of modified kernel would have extra two powers of $G$ in the denominator, while the coordinate representation will be non-divergent. 

As a second possibility, one can expand the unknown current over the orthogonal basis functions $\phi_n(x)$ that already satisfy the zero boundary condition:
\begin{equation}
    j_s(x) = \sum_n{c_n \phi_n(x)},\qquad \phi(0) = \phi(W) = 0.
\end{equation}
The resulting matrix equation
\begin{equation}
    \left[Z_{nm} - \frac{\delta_{nm}}{\sigma_{\rm m}}\right]c_m = (\mathcal{E}_0)_n
\end{equation}
would have matrix elements $Z_{nm}$ that quickly converge at large $G$. 

We have chosen the basis functions
\begin{equation}
    \phi_n(x) = \sqrt{\frac{2a}{W}}\sin\left(\frac{\pi n a}{W}\right)
\end{equation}
that are orthogonal with respect to the inner product:
\begin{equation}
    \int_0^W{\frac{dx}{a}\phi_n(x)\phi_m(x)} = \delta_{nm}.
\end{equation}
In this basis, the elements of the impedance matrix are evaluated analytically at each $G$:
\begin{equation}
   (Z_G)_{nm} = 2 \pi ^2 m n \frac{W}{a}\frac{ \left((-1)^m e^{-i G W} - 1 \right) \left((-1)^n e^{i G
   W} - 1\right)}{\left(\pi ^2 m^2-G^2 W^2\right) \left(\pi ^2 n^2-G^2 W^2\right)} Z_G.
\end{equation}
Expansion of current density over Chebyshev polynomials $U_n(x)$ multiplied by their respective weight functions $w(x)$ is advantageous in predicting the character of the field at the edges. However, it comes at the cost of numerical approximation to matrix elements $Z_{nm}$.

In actual calculations, we have truncated the linear system at $n = m = 10$, and evaluated the sums up to $G_{\max} = 20 (2 \pi /a)$. 



\end{widetext}

\end{document}